\documentclass[twocolumn,showpacs,preprintnumbers,amsmath,amssymb]{revtex4}
\usepackage{graphicx}
\usepackage{dcolumn}
\usepackage{bm}
\def\ph2{{\it p}-H$_2$}
\def\gapx{\lower 2pt \hbox{$\buildrel>\over{\scriptstyle{\sim}}$\ }}
\def\lapx{\lower 2pt \hbox{$\buildrel<\over{\scriptstyle{\sim}}$\ }}
\def\Am3{\AA$^{-3}$}
\begin{document}
\title{Superfluidity and Quantum Melting of \ph2 Clusters}

\author{Fabio Mezzacapo}
\author{Massimo  Boninsegni}

\affiliation{Department of Physics, University of Alberta, Edmonton, Alberta, Canada T6G 2J1 }
\date{\today}

\begin{abstract}
Structural and superfluid properties of \ph2 clusters of size up to $N$=40 molecules, are studied at low temperature (0.5 K $\le$ $T$ $\le$ 4 K) by path integral Monte Carlo simulations. The superfluid fraction $\rho_S(T)$ displays an interesting,  non-monotonic behavior  for $22\le N\le30$. We interpret this dependence in terms of  variations with $N$ of the cluster  structure. Superfluidity is observed at low $T$ in clusters of as many as 27 molecules; in the temperature range considered here, {\it quantum melting}  is observed in some clusters, which are seen to freeze at high temperature.
\end{abstract}

\pacs{67.90.+z, 61.25.Em}

\maketitle
Recent developments in spectroscopy afford the investigation of properties of a single complex  molecule embedded in clusters of $^4$He or \ph2.  Specifically, by studying the rotational spectrum of the molecule one can obtain evidence of decoupling  of its rotation from the surrounding medium (i.e., the cluster), at  sufficiently low temperature (of the order of a fraction of 1 K). Such a decoupling is interpreted as due to onset of superfluidity (SF) in the cluster \cite{grebenev00}. 
This is now an area of intense, current research effort, aimed at gaining theoretical understanding of the microscopic origin of SF, perhaps the most fascinating manifestation of quantum behavior on a macroscopic scale. In particular, theoretical questions are being addressed such as: {What is the smallest finite size system for which SF can be observed ?} 
{Which {\it condensed} (i.e. non-gaseous) matter systems, besides helium, can display this phenomenon, if not in the bulk at least in sufficiently small clusters ?}

With respect to the second question,  droplets of hydrogen molecules are clearly of fundamental interest. Molecular {\it para}-Hydrogen has long been speculated to be a potential superfluid, owing to the bosonic character and the light  mass of its constituents \cite{ginzburg72}. However, the search for SF in bulk \ph2 has so far been frustrated by the fact that, unlike helium, this system solidifies at low temperature, as  the intermolecular potential is significantly more attractive than that between two helium atoms. 
On the other hand, clusters of \ph2 molecules of sufficiently small size, ought to  remain ``liquidlike''  at significantly lower temperature than the bulk, possibly turning superfluid \cite{note}.

Indeed, theoretical studies carried out some fifteen years ago, based on path integral Monte Carlo (PIMC) simulations \cite{sindzingre91}, yielded evidence of a finite superfluid response in pure  [($p$-H$_2$)$_N]$ clusters, with $N$=13 and $N$=18  molecules, at a temperature $T$ $\lapx$2 K, whereas a larger cluster ($N$=33) was found to be ``solidlike", nonsuperfluid, in the same temperature range. Other PIMC calculations showed that, while SF may not necessarily occur in larger clusters, there is nonetheless a large propensity for quantum exchanges \cite{scharf92}. Finally, more recent PIMC studies have yielded evidence of superfluid behavior in small (17 molecules) clusters of \ph2 doped with a single OCS \cite{kwon02,W2} or CO \cite{Saverio} molecule. 
 
For pristine \ph2 clusters, the only theoretical  study of superfluidity  is restricted  to three  cluster sizes ($N$=13, 18, and 33) \cite{sindzingre91}. No systematic study of the superfluid properties of clusters as a function of size and 
temperature has yet been carried out. Although experimental data are not yet available for pure \ph2 clusters, novel techniques based on Raman spectroscopy hold promise for the investigation of superfluidity in these systems \cite{toennies}. 

In this Letter,  we present a detailed investigation of  ($p$-H$_2$)$_N$ clusters at low temperature (down to $T$=0.5 K), for  $N$ $\le$ 40, by means of PIMC simulations, based on a recently developed {\it worm} algorithm \cite{MBworm}. This numerical technique affords accurate estimates of thermodynamic properties of Bose systems. In particular, the superfluid fraction can be calculated  with much  greater accuracy than that afforded by conventional PIMC, which has been one of the leading many-body computational methods of the last 20 years \cite{ceperley95}. 

Our main findings are the following:  (\ph2)$_N$ clusters with $N<22$ are liquidlike, and superfluid at low $T$. Superfluid properties of clusters with 22 $\le$ $N$ $\le$ 30,  strongly depend on  $N$.  
A few clusters, in the range $22 \le N \le 30$,  feature, in temperature range considered here, a behavior that is consistent with coexistence of  insulating (solidlike) and superfluid (liquidlike) ``phases'', the latter becoming prominent as $T$ is lowered. In other words, such clusters melt at low $T$ as a result of  zero-point motion, and freeze at higher temperature. We refer to this intriguing behavior as ``quantum melting".  The superfluid response of clusters with $N\ge30$ is significantly depressed;  notably, however, for $N=40$ permutation cycles can still be observed including  a considerable number (as many as 20) of \ph2 molecules. 

We model our system of interest as a collection of $N$  \ph2 molecules, regarded as point particles and   interacting via an accepted pair potential \cite{SilveraandGoldman}.  The system is assumed to be at a temperature $T$=1/$\beta$.
Since we are interested in studying properties of the clusters as a function of $N$, we use a variant of the worm algorithm described in Ref. \cite{MBworm}, in which the number of particles in the configurations inside the so-called $Z$ sector (those that contribute to the expectation values of physical observables) is fixed at $N$. We utilized a high-temperature approximation for the many-body density matrix accurate up to fourth order in the imaginary time step $\varepsilon$ \cite{chin}, and used for our calculations a value of the imaginary time step $\varepsilon=1/640$ K$^{-1}$, which we empirically found to yield converged estimates. 
We computed cluster energetics,   radial density profiles and the  superfluid fraction $\rho_S(T)$ (using the well-known ``area" estimator \cite{sindzingre91}).

\begin{figure}
\centerline{\includegraphics[scale=0.32, angle=-90]{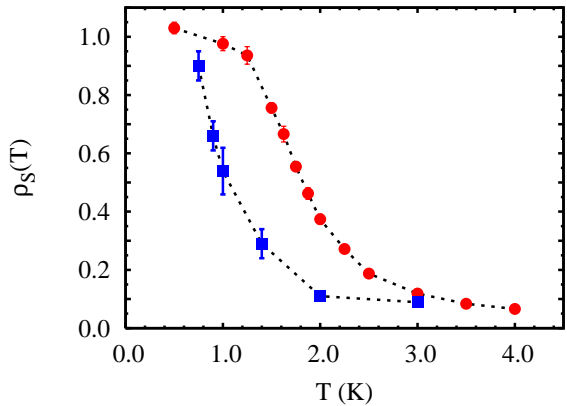}}
\caption{(color online) Superfluid fraction $\rho_S(T)$ for  clusters of 20 (circles) and 23 (boxes) \ph2 molecules. Dotted lines are guides to the eye. When not shown, statistical errors are smaller than the symbol size. }
\label{fig:20t}
\end{figure}

Figure \ref{fig:20t} shows  $\rho_S(T)$ for clusters of size $N$=20 and $N$=23.  In both cases, as expected  $\rho_S(T)$ is a monotonically decreasing function of $T$. As we discuss below and later in the manuscript, however, the physical behavior of these two clusters is qualitatively different.  

For $N$=20, the behavior of $\rho_S(T)$ is  close to that observed in Ref. \onlinecite{sindzingre91} for $N$=18. At $T\le$ 1.25 K, the system is essentially entirely  superfluid;  $\rho_S(T)$  drops rather quickly to a value $\simeq 0.2$ at $T$=2.5 K (corresponding to roughly four molecules in the superfluid phase), and decreases more slowly at  higher temperatures. While there is obviously no real phase transition in a finite system, the notion of  ``superfluid fraction" becomes scarcely meaningful, when the average number of molecules in the superfluid phase  is of the order of 1.  Therefore, we operatively define our ``transition temperature" $T_{\rm c}$ as that at which $N\rho_S(T_{\rm c})\sim 2$. For $N$=20, this heuristic criterion yields $T_{\rm c}\sim$ 3 K. Remarkably, however,  even at this temperature the probability for a \ph2 molecule to belong to a permutation cycle involving three or more molecules is still as large as $\sim$ 3\%, and exchange cycles involving as many as 13 molecules are observed. This is consistent with the qualitative observation made in Ref. \onlinecite{scharf92}, based on a PIMC simulation which did not explicitly include exchanges.  
\begin{figure}
\centerline{\includegraphics[scale=0.32, angle=-90]{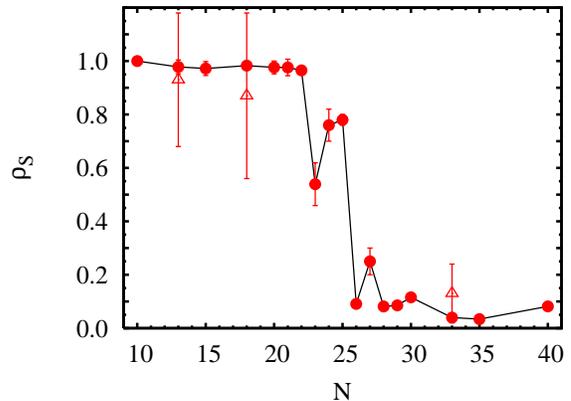}}
\caption{(color online) Superfluid fraction versus cluster size $N$, at $T$=1 K (filled circles). When not shown, statistical errors are of the order of, or smaller than the symbol size. Solid line is only a guide to the eye. Also shown for comparison are results from Ref. \onlinecite{sindzingre91} (open triangles).}
\label{fig:1}
\end{figure}

As the number $N$ of particles increases, the physics of a cluster ought to approach that of the bulk; whereas this means a liquid (superfluid at low 
$T$) for helium clusters,  \ph2 forms an insulating crystal. It is therefore reasonable to expect that, at some fixed, low $T$ (e.g., $T$=1 K),  the superfluid fraction of \ph2 clusters should decay to zero as $N\to\infty$.

Figure \ref{fig:1} shows our estimates for the superfluid fraction $\rho_S$ as a function of the cluster size $N$, at $T$=1 K. For $N\le 22$, (\ph2)$_N$ clusters are entirely superfluid (within the precision of our calculation) i.e.,   $\rho_S$ $\approx$ 1.  Furthermore, the temperature dependence of $\rho_S$ is consistently similar to that shown in Fig. \ref{fig:20t} (for $N$=20). 
In the range $22\le N \le 30$ a remarkable, nonmonotonic behavior of $\rho_S$ is observed, with dramatic differences between clusters differing by the addition of just one molecule.  
We propose that such an interesting behavior reflects alternating liquidlike (superfluid) or solidlike (insulating) character of the clusters, strongly dependent on $N$.    This interpretation is consistent with the expectation that a crystalline (nonsuperfluid) phase should  emerge, at large $N$.

For example, at $N$=25 the cluster displays a large  superfluid response at $T$=1 K. If a single molecule is added, the superfluid fraction drops abruptly, to a value less than 0.1. On adding yet another molecule, the superfluid fraction grows again, to approximately 0.25. It seems difficult to imagine that the addition of a single molecule would alter so drastically the superfluid component, if the cluster structure stayed essentially the same, e.g., liquidlike.  It seems reasonable, instead, to relate the changes in the superfluid properties to structural changes that occur on adding molecules.

\begin{figure}
\centerline{\includegraphics[scale=0.32, angle=-90]{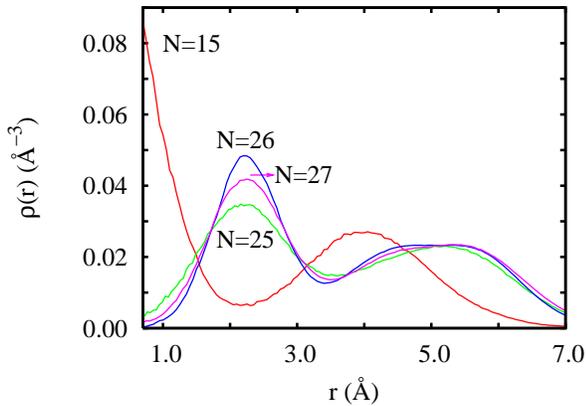}}
\caption{(color online) Radial density  computed with respect to the center of mass, for clusters with 15, 25, 26, and 27 \ph2 molecules. Statistical errors, not shown for clarity, are of the order of 5$\times10^{-4}$ \Am3 or less.  }
\label{fig:rhorad}
\end{figure}

In order to illustrate this point, we show in Fig. \ref{fig:rhorad} profiles of  radial density  ($\rho(r)$), computed with respect to the center of mass of the cluster, at $T$=1 K. For $N$=15, the large value of $\rho(r\to 0)$ and the local minimum for $r\approx$ 2 \AA\ indicate the presence of a  single \ph2 molecule  in the center of the cluster. Other molecules form a floppy shell around the central one, as shown by the peak at $r$=4 \AA. On increasing the cluster size,  qualitative changes occur at $N$ $\sim$ 22.  The value of $\rho(r\to 0)$ becomes negligible, i.e.,  the center of the cluster is no longer occupied by a molecule. There is a peak at about 2 \AA\ from the center, as an inner molecular shell forms. A second, broader peak at larger distance ($r\approx$ 5 \AA) corresponds to the formation of an outer shell.  

The main structural change, going from $N$=25 to $N$=26, is that the first peak becomes significantly sharper, and its height increases by some 40\% (see Fig. \ref{fig:rhorad}).  We interpret this as evidence that  the inner shell becomes more solidlike, with molecules localized and  quantum exchanges depressed, both in the first shell as well as between the first and second shells. If another molecule is added, the density profile for $N$=27 features a first-shell peak and an intershell minimum of heights intermediate between those of the $N$=25 and $N$=26 cases, and  $\rho_S$ increases to a value  much lower than for $N$=25, but significantly greater than that for $N$=26.  Thus, the addition of a molecule to the $N$=26 has the effect of frustrating the solid order of the inner shell, increasing molecule delocalization and leading to quantum exchanges.

\begin{figure}
\centerline{\includegraphics[scale=0.32, angle=-90]{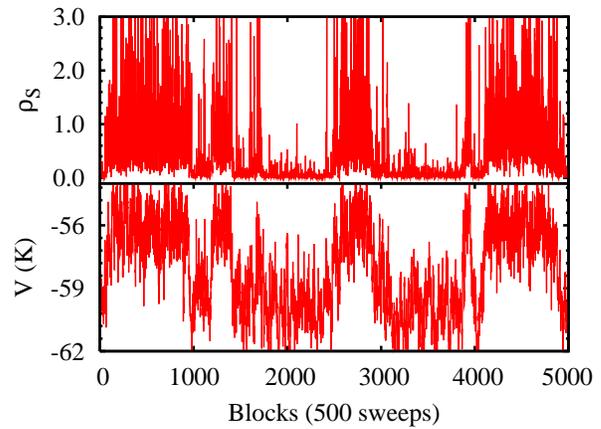}}
\caption{(color online) Behavior of  superfluid fraction (upper panel) and potential energy per molecule (lower panel) observed during a typical Monte Carlo run for a cluster of $N$=23 molecules at $T$=1 K. Data shown refer to successive block averages of $\rho_S$ and $V$. A single ``block" consists of 500 sweeps through the entire system (see text). Although large fluctuations are present, visual identification is relatively easy of  two different regimes, in which $\rho_S$ is on average either close to 1 or zero; correspondingly $V$ oscillates between two values close to $\sim - 55$ K (liquid) and $\sim - 60$ K (solid) respectively.  }
\label{fig:area}
\end{figure}

\begin{figure}
\centerline{\includegraphics[scale=0.32, angle=-90]{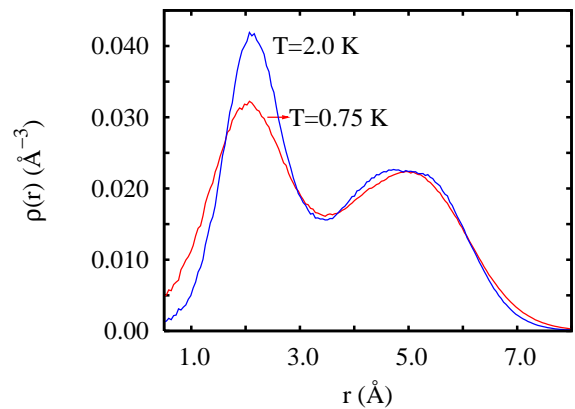}}
\caption{(color online) Radial density profiles computed for a cluster of $N$=23 molecules at $T$=2.0 K and $T$=0.75 K. Statistical errors, not shown for clarity, are of the order of 5$\times10^{-4}$ \Am3 or less.}
\label{rt}
\end{figure}
A particularly intriguing behavior is observed in some clusters; we discuss in detail the  $N$=23 one, for which $\rho_S(N)$ at $T$=1 K takes on a local minimum. Fig. \ref{fig:area} shows  the values of the superfluid fraction (upper panel) as well as of the potential energy per particle ($V$, lower panel), recorded in a typical Monte Carlo run (results shown in the figure correspond to a few hundred hours of CPU time on a high-end workstation). Specifically, data shown refer to successive block averages of $\rho_S$ and $V$, each block consisting of 500 sweeps through the system \cite{notex}. 
The quantity $\rho_S$ displays large fluctuations, but generally switches abruptly between two clearly identifiable regimes, namely, one in which it oscillates around an average value of one, and another in which it stays  close to zero. In close correspondence with $\rho_S$, the potential energy $V$ displays similar oscillations, switching between configurations of lower (for small $\rho_S$) and higher (greater $\rho_S$) values.

We interpret  this pattern as the signature of the coexistence of two cluster phases, characterized by large and small superfluid response.  The behavior  of the potential energy suggests  that, in its non-superfluid phase, a cluster comprising 23 \ph2 molecules should feature distinct solidlike properties, chiefly a high degree of localization of the molecules. The   coexistence of these two (solid- and liquidlike) phases renders the precise determination of the average value of $\rho_S$ computationally rather demanding, i.e., fairly lengthy runs are needed. On decreasing the temperature, the liquidlike superfluid phase becomes dominant, i.e., the cluster ``melts" at low $T$ due to quantum zero-point motion of the \ph2 molecules, ``freezing" instead at higher temperature.  This is consistent with the observed evolution of the radial density profile, shown in Fig. \ref{rt} for $T$=2.0 K and 0.75 K. As $T$ is lowered, the first peak broadens significantly, as molecules enjoy greater mobility.

Qualitatively similar results are seen for other clusters, e.g., $N$=27, in the temperature range explored in this work. It is likely that other clusters, in the range $22 \le N\le 30$, may display the same behavior, at some temperature (in some cases possibly much lower than $T$=0.5 K, which is the lowest considered here). 
It is important to note that this behavior is markedly different than that observed in clusters with $N <$ 22, for which a plot such as that of Fig. \ref{fig:area} merely shows the two quantities $\rho_S$ and $V$ fluctuate around their average values, with no evidence of   the system switching back and forth between two distinct phases.
Indeed, clusters with $N$$<$ 22 are found to be liquidlike at all temperatures, with a growing superfluid (normal) component at low (high)  $T$.  

Summarizing, we have studied  superfluid and structural properties of \ph2 clusters of  size $N \le 40$.  We observed  nontrivial superfluid behavior of  (\ph2)$_N$, as a function of $N$.  Our observation is consistent  with the emergence of a solid phase, as the size of the cluster grows; however, this occurs non-monotonically. Some clusters (e.g., $N$=23 and 27) 
feature, at low temperature, quantum melting, induced by zero-point motion; these clusters are observed to  freeze at high temperature.
Some of these predictions may soon be tested experimentally \cite{toennies}.  

This work was supported  by the Natural Science and Engineering Research Council of Canada under research grant 121210893.   


\begin{thebibliography}{9}
\bibitem{notex}
A sweep is defined as a sampling cycle that begins with the opening of a worm and ends with its reconnection; see Ref. \onlinecite{MBworm} for details.
\end{thebibliography}
\end{document}